\documentclass[twocolumn]{aastex63}
\usepackage{amsmath}
\usepackage{enumitem}
\usepackage{ulem}
\submitjournal{ApJ}

\shorttitle{Identifying and Repairing Catastrophic Errors in Galaxy Properties}
\shortauthors{Hovis-Afflerbach et al.}

\begin{document}

\title{Identifying and Repairing Catastrophic Errors in Galaxy Properties Using Dimensionality Reduction}

\author[0000-0002-9967-2725]{Beryl Hovis-Afflerbach}
\affiliation{California Institute of Technology, 1200 E California Blvd, Pasadena, CA 91125, USA}
\affiliation{Cosmic Dawn Center (DAWN)}
\affiliation{Niels Bohr Institute, University of Copenhagen, Lyngbyvej 2, DK-2100 Copenhagen \O }
\author[0000-0003-3780-6801]{Charles L. Steinhardt}
\affiliation{Cosmic Dawn Center (DAWN)}
\affiliation{Niels Bohr Institute, University of Copenhagen, Lyngbyvej 2, DK-2100 Copenhagen \O }
\author[0000-0001-5382-6138]{Daniel Masters}
\affiliation{IPAC, Caltech, 1200 E California Blvd, Pasadena, CA 91125}
\author[0000-0001-7116-9303]{Mara Salvato}
\affiliation{Max Planck Institute for Extraterrestrial Physics (MPE) \& Excellence Cluster Universe. Giessenbachstr. 1, D-85748 Garching. Germany}

\begin{abstract}
Our understanding of galaxy evolution is derived from large surveys designed to maximize efficiency by only observing the minimum amount needed to infer properties for a typical galaxy. However, for a few percent of galaxies in every survey, these observations are insufficient and derived properties can be catastrophically wrong. Further, it is currently difficult or impossible to determine which objects have failed, so that these contaminate every study of galaxy properties. We develop a novel method to identify these objects by combining the astronomical codes which infer galaxy properties with the dimensionality reduction algorithm t-SNE, which groups similar objects to determine which inferred properties are out of place. This method provides an improvement for the COSMOS catalog, which already uses existing techniques for catastrophic error removal, and therefore should improve the quality of large catalogs and any studies which are sensitive to large redshift errors.
\end{abstract}

\keywords{astronomy data analysis --- redshift surveys --- high-redshift galaxies --- photometry --- dimensionality reduction --- galaxy properties}



\section{Introduction}

Many recent advances in galaxy evolution have only been possible with large photometric surveys, in which objects are imaged in a small number of typically broad filters. These observations can then be fit with with increasingly detailed templates derived from a combination of physical models and high-resolution observations to determine the properties of each object.  Many discoveries have been made using these best fit properties, including those of the star-forming `main sequence' (cf. \citealt{Noeske2007,Peng2010,Speagle2014}) and high-redshift quiescent galaxies \citep{Toft2007,Ilbert2013,Glazebrook2017}.

Although photometric template fitting is highly successful at inferring the properties of most galaxies, it also has an unusual failure mode.  With the limited information provided by photometry, features such as the Balmer and Lyman breaks are easily confused, so that many probability density functions have multiple peaks.  In a small number of cases, the best-fit template will lie near the wrong peak, producing properties which are catastrophically incorrect.  A comparison between the results of many existing template fitting algorithms and spectroscopic redshifts ($z_{spec}$) found these `catastrophic errors' in redshift, defined as $\frac{\Delta z}{1+z_{spec}} > 0.15$, occur in at least 5\% of objects \citep{Hildebrandt_2010}.

\begin{figure*}
    \centering
    \includegraphics[width=.9\textwidth]{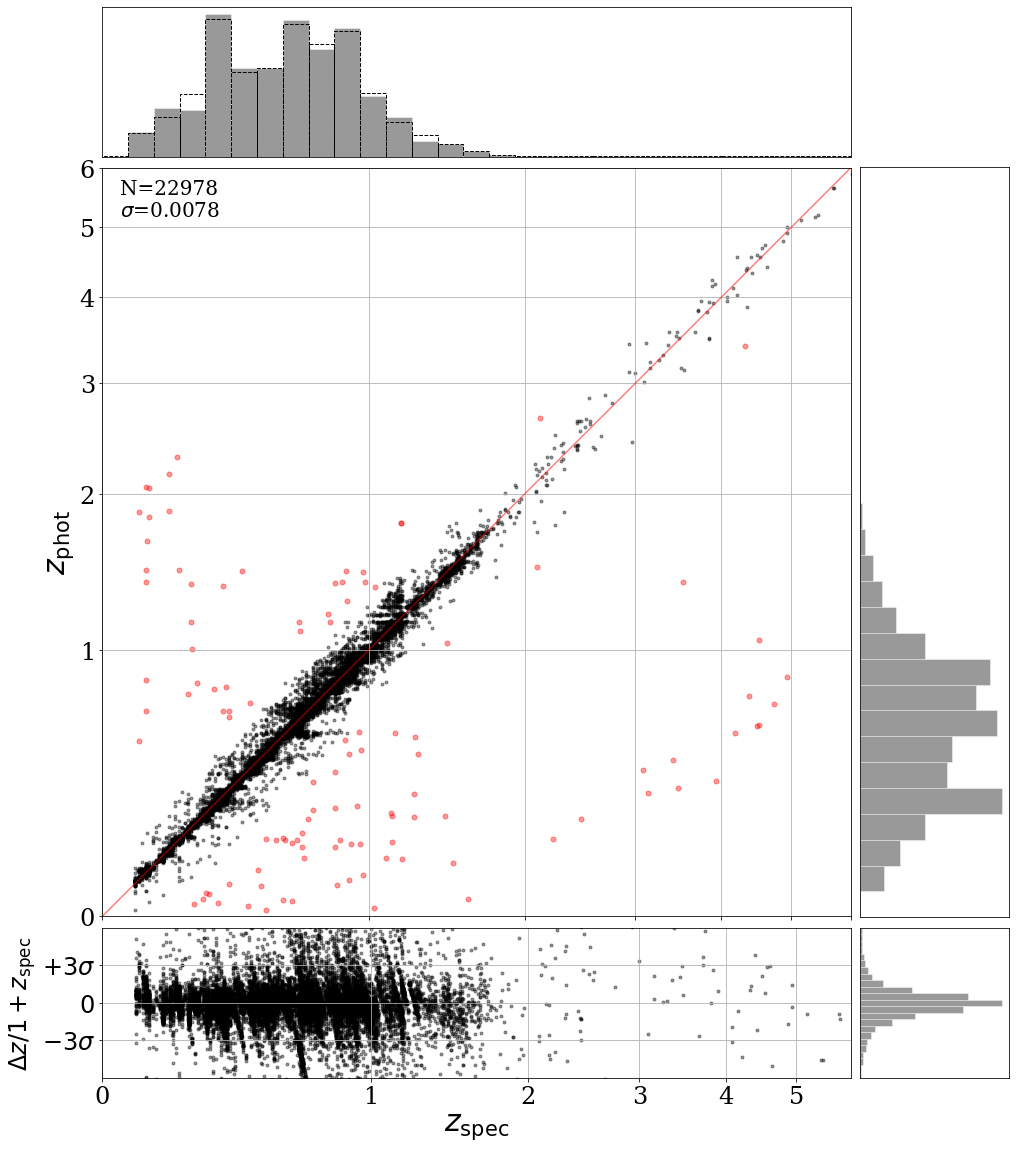}
    \caption{Comparison of photometric redshift with spectroscopic redshift for a sample of 22,978 objects from the COSMOS survey \citep{Laigle_2016, Jin_2018}, with z$_{phot}$ determined using EAZY \citep{Brammer_2008}. 99.5\% of objects (in black) fall on or near the x=y line, while the other 0.5\% of objects (catastrophic errors, in red) fall greater than $\frac{\Delta z}{1+z_{spec}} > 0.15$ away from this line, a threshold defined in previous studies \citep[e.g.][]{Hildebrandt_2010}.
    Below this comparison is a distribution of error around the $x=y$ line. The (red) objects with catastrophic errors have errors much greater than $\pm 6 \sigma$, so they do not appear on this plot.
    In the bottom panel, the normalized median absolute deviation $\sigma_{NMAD} = 1.48 \times \mathrm{median} \left( \left| \frac{\Delta z - \mathrm{median} (\Delta z)}{1 + z_{spec}} \right| \right)$ \citep{Brammer_2008} is chosen to be less sensitive to outliers or catastrophic errors.
    The three histograms show the relative distribution of objects in the sample within each redshift bin.  The $z_{phot}$ distribution (dashed) is overlaid on the $z_{spec}$ distribution at top for comparison. The horizontal bands at constant $z_{phot}$ are due to the coarseness of the redshift grid and gaps between bands. However, they do not contribute to the catastrophic errors examined in this work.}
    \label{fig:zspeczphot}
\end{figure*}

Even this low catastrophic error rate has significant consequences.  Because high-redshift objects are rare, a significant fraction of objects with high photometric redshift ($z_{phot}$) truly lie at lower redshift (Fig. \ref{fig:zspeczphot}).  In SPLASH, nearly half of all objects with $z_{phot} > 4$ and existing spectra were found to have $z_{spec} < 2$ \citep{Steinhardt2014}.  Followup observations find that a similar fraction of high-redshift quiescent galaxy candidates lie at lower redshift instead.

These catastrophic errors likely come from situations in which templates cannot sufficiently describe the observed photometry.  This may be due to an actual difference between the galaxy spectrum and templates from physical models, or instead due to near-degeneracies combined with measurement uncertainty.  As a result, approaches based on modeling different astrophysics have also been considered. For example, some studies have used clustering information to infer redshift information from the angular distribution of sources within the catalog \citep{Hildebrandt2009,McQuinn_2013,Rahman_2015,Morrison_2017}.

Recently, alternative approaches based on machine learning algorithms have been investigated \citep{Masters_2015,Speagle_2017,Hemmati_2019}.  These methods take the opposite approach to template fitting and clustering by avoiding model dependence.  Assuming objects with similar photometric colors have similar properties, no physical model is necessary to infer those properties.  Rather, it is only necessary that each galaxy with an uncertain redshift has similar colors to at least one object which does have well-measured properties or a followup spectrum.

Here, we consider whether an augmented approach is possible in which no followup spectroscopy is required, but instead dimensionality reduction is used to correct catastrophic redshift errors solely through comparison with other photometric redshifts.
Such an approach is also used in classification-aided photometric redshift estimation \citep[CPz,][]{Fotopoulou_2018}, which uses a random forest algorithm.

In \S~\ref{sec:methods}, we describe the construction of a map based on t-distributed Stochastic Neighbor Embedding (t-SNE;  \citealt{vanDerMaaten_2008,vanDerMaaten_2014}) such that close neighbors have similar colors. In \S~\ref{sec:results}, we create this map using the COSMOS2015 catalog \citep{Laigle_2016} and the COSMOS superdeblended catalog \citep{Jin_2018}. We show that many catastrophic redshift errors will appear as outliers when that map is colored by $z_{phot}$, enabling the identification of these errors without spectroscopy, and consider whether this can be used to not only identify, but also repair these catastrophic errors.  Finally, the implications of this experiment for future photometric catalogs are discussed in \S~\ref{sec:discussion}.

\section{Developing an Augmented Algorithm}
\label{sec:methods}

Photometric template fitting is highly successful, but treats each object individually according to physical models.  The goal of this work is to take advantage of information about objects with similar colors to correct the small fraction of catastrophic redshift errors, without the need for a training sample of objects with spectroscopically confirmed redshifts.

The approach taken here relies on two plausible assumptions: (1) that photometric template fitting is correct for most objects; and (2) that photometric bands are chosen such that objects with similar colors should lie at similar redshifts (and, presumably, have similar physical properties in general).  Then, it should be possible to produce improved redshifts as follows:
\begin{enumerate}
    \item Run a standard photometric template fitting code on the entire sample.  In this work, we choose the code EAZY (\citealt{Brammer_2008}; see \S~\ref{subsec:eazy}).
    \item Use t-distributed Stochastic Neighbor Embedding (t-SNE; see \S~\ref{subsec:tsne}) to group every object in the catalog such that each is near objects with similar photometric colors.
    \item For every object in the catalog, consider whether its $z_{phot}$ is similar to that of its neighbors.  If so, accept the redshift, and otherwise flag it as a likely error.
    \item For objects flagged as likely errors, use the redshifts of objects with similar photometry to produce a new redshift estimate.
\end{enumerate}

This algorithm is evaluated using a sample of 23,039 galaxies with photometry from the COSMOS2015 catalog \citep{Laigle_2016} matched with spectroscopic redshifts from the COSMOS superdeblended catalog \citep{Jin_2018}.  Spectroscopic redshifts are assumed to be the ground truth.  Thus, successful identification of catastrophic redshift errors consists of identifying a sample where $z_{spec}$ and $z_{phot}$ disagree substantially, and successful repair of these errors consists of producing an updated $z_{phot}$ which agrees with the measured $z_{spec}$.  

The reduction to a spectroscopic sample is necessary because tests can only be performed using a sample with known "true" redshifts.  However, it should be noted that this produces a test sample which is generally brighter and better-measured than the full COSMOS2015 catalog.  The problem considered here is therefore easier than it would be for the full catalog, and potentially also restricted to a smaller region of color space (\citealt{Masters_2015}; Fig. 6).  One consequence is that in a relative sense, the false positive rate of a detection algorithm will be more significant on the spectroscopic catalog, whereas the false negative rate will be more significant on the full catalog.

\subsection{EAZY}
\label{subsec:eazy}
We use the Easy and Accurate Redshifts from Yale (EAZY, \citealt{Brammer_2008}) template fitting code to determine $z_{phot}$ values for the sample.
We choose this code because it is a standard code used for template fitting, and because it performs relatively well and produced a typical but slightly lower outlier ($\frac{\Delta z}{1 + z_{spec}} > 0.15$) fraction in a comparison with other template fitting codes \citep{Hildebrandt_2010}.

EAZY works by fitting a set of templates to the photometry for each input galaxy. To determine a galaxy's properties, it chooses the linear combination of templates which minimizes the fit error.  EAZY's method is different than that of a number of other codes in that it uses a basis of templates, rather than on a grid.  Some of the catastrophic errors that arise in running a grid-based code are often caused by the grid being too sparse, and for a number of objects the code fails to find the correct minimum.  By contrast, the errors that arise in running EAZY are more likely to arise because the template set is limited, leading to degeneracies between potential spectra at different redshift values.

For this work, EAZY was run with the F160 prior and the \texttt{tweak\_fsps\_QSF\_12\_v3} template set.
Input data includes 32 bands from the COSMOS 2015 catalog for each object in the spectros sample:
\setlist{nolistsep}
\begin{itemize}[noitemsep]
    \item Subaru Hyper-Supreme-Cam $Y$ band
    \item UltraVISTA NIR $Y, \ J, \ H, \ K_s$ bands
    \item CFHT WIRCam $K_s, \ H$ bands
    \item CFHT MegaCam $u$ band
    \item Subaru Suprime-Cam broad bands $B, \ V, \ r, \ i^+$, $z^{++}$
    \item Subaru Supreme-Cam medium bands $IA427$, $IA464, \ IA484, \ IA505, \ IA527, \ IA574, \ IA624$, $IA679, \ IA709, \ IA738, \ IA767, \ IA827$
    \item Subaru Supreme-Cam narrow bands $NB711$, $NB816$
    \item Spitzer IRAC $ch1,\ ch2,\ ch3,\ ch4$ from the SPLASH survey
    \item GALEX $NUV$ band
\end{itemize}
The sample was restricted to objects flagged as galaxies (\texttt{type=0}).  Not every object has coverage in every band, as summarized in \cite{Laigle_2016}. This poses no issues for EAZY, however to run t-SNE it requires us to later cut the catalog to eight bands (see \S~\ref{subsec:tsne}) and the 1,061,787 objects (89.8\% of the catalog) which have coverage and are not saturated in those 8 bands. Additionally, we use spectroscopic redshifts from \citet{Jin_2018} as ground truth, cutting the sample to 23,039 objects. We were able to determine photometric redshifts using EAZY for 22,978 of these. For the other 61 objects, the EAZY models did not converge and EAZY was not able to assign a $z_{phot}$, which typically occurs for a small number of objects.

A comparison between the photometric redshifts determined by EAZY with the spectroscopic redshifts shows that most are in strong agreement, while a small fraction of the objects have catastrophic errors (Fig.  \ref{fig:zspeczphot}).

\begin{figure}
    \centering
    \includegraphics[width=0.47\textwidth]{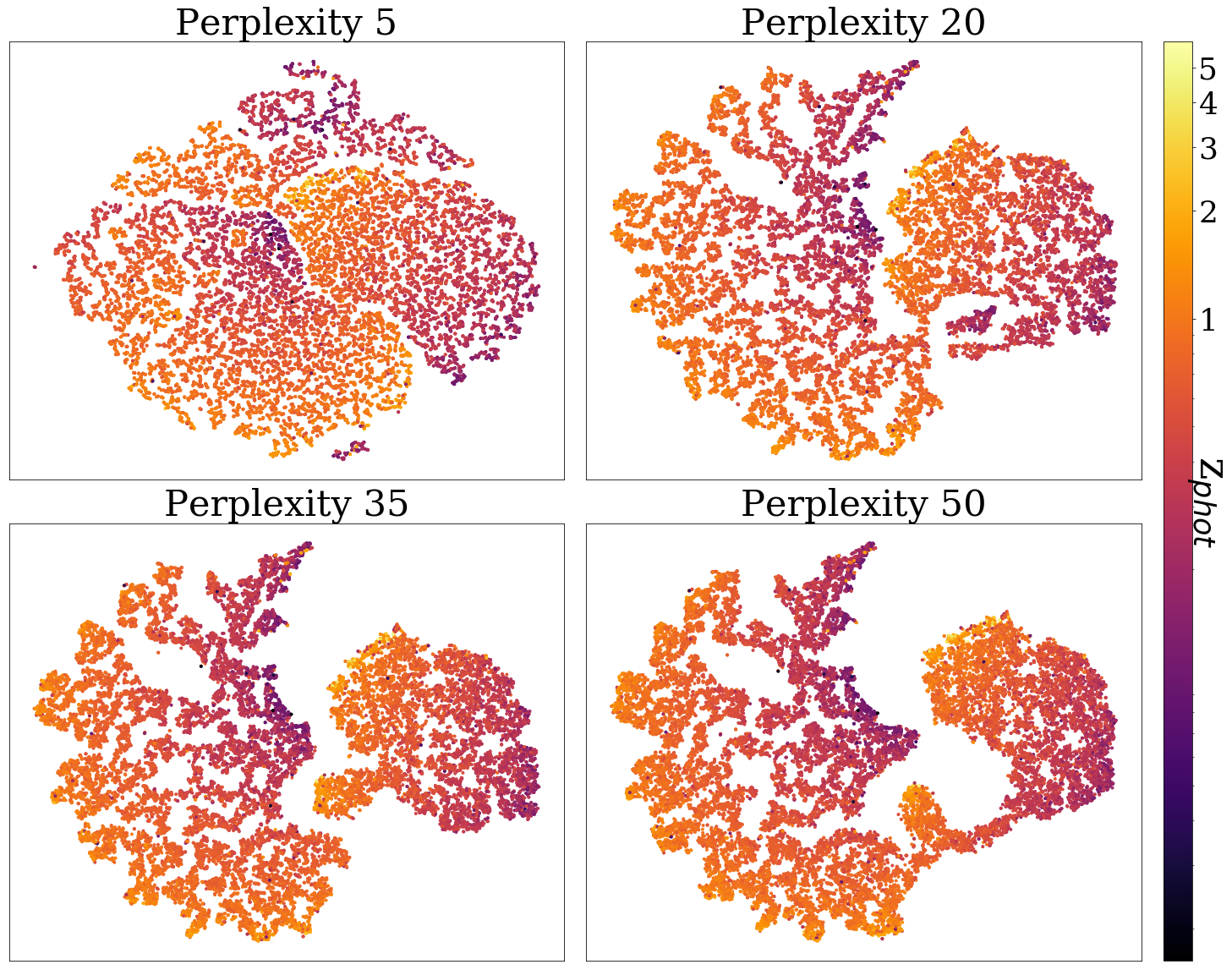}
    \caption{t-SNE output for all objects with 5000 iterations and perplexities 5, 20, 35, and 50, spanning the range recommended in \citet{vanDerMaaten_2008}. 
    Higher values of perplexity result in greater global structure and more clustering.
    Each point, each galaxy's representation in the two-dimensional map, is colored according to the galaxy's $z_{phot}$.}
    \label{fig:perplexity}
\end{figure}

\subsection{t-SNE}
\label{subsec:tsne}

Independent of the EAZY fitting, the next step is to select a set of galaxies with similar photometry for every galaxy in the catalog.  Ideally, this selection would have been done in the full (eight-dimensional in this case) color space.  However, even a catalog with 23,000 galaxies, as used here, is sparse in eight dimensions, and would rapidly become sparser as more bands are used.  This is one example of a problem known as the ``curse of dimensionality'' \citep{Bellman_1961}, and is the reason that a dimensionality reduction algorithm is run prior to selecting neighbors, with the goal of producing a denser, lower-dimensional space that preserves as much of the structure of the higher-dimensional space as possible.

Here, the algorithm selected is t-distributed Stochastic Neighbor Embedding (t-SNE; \citealt{vanDerMaaten_2008, vanDerMaaten_2014}). t-SNE is an unsupervised algorithm which has been used in similar ways to identify quiescent galaxies from their photometry \citep{Steinhardt_2020} and classify gamma-ray bursts \citep{Jespersen_2020}.  The goal is to map galaxies into a two-dimensional space such that galaxies with similar photometry are placed as close neighbors and galaxies with dissimilar photometry are not.

The resulting t-SNE map depends heavily on additional choices made by the user, both in preprocessing the data and in determining the relative importance of local and global structure.  Specifically, in this work the following procedure is used:

\begin{figure*}[ht]
    \centering
    \includegraphics[width=\textwidth]{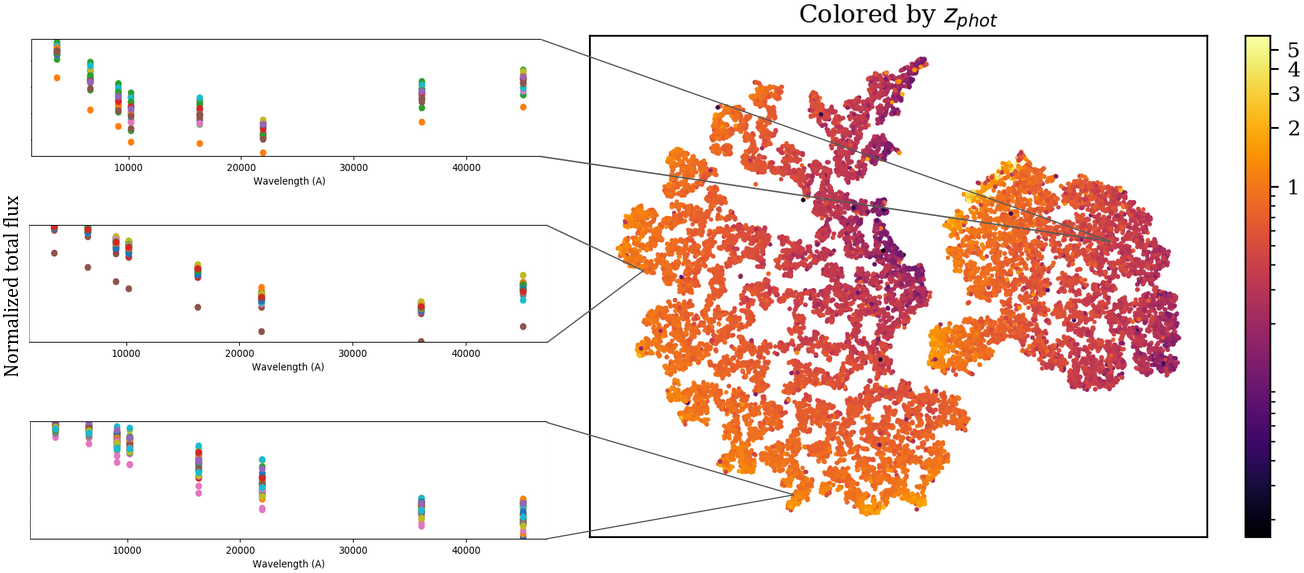}
    \caption{t-SNE output map with 5000 iterations and perplexity 35, colored by $z_{phot}$.
    For a handful of objects in each of three distinct regions in the low dimensional t-SNE output map, the high dimensional input data, the SED shape, is shown in the boxes at left.
    Objects grouped close together on the t-SNE map, which are in the same box at left, have similar SED shapes.
    By contrast, the SED shapes are different between these three distinct regions of the t-SNE map.}
    \label{fig:tsnesed}
\end{figure*}

\begin{enumerate}
    \item Input data consists of normalized total fluxes for each object at eight bands (u, r, z$^{++}$, yHSC, H$_w$, Ks$_w$, IRAC1, IRAC2), which due to normalization lie in a seven-dimensional space.  This is intended to describe the shape of each galaxy's spectral energy distribution (SED) without using the overall magnitude to predict redshift.  Since t-SNE cannot handle missing or incomplete data, only galaxies with measured fluxes in all eight bands are included; galaxies with either a lack of coverage in COSMOS2015 or a non-detection in any band are discarded. These eight bands are chosen to comprehensively describe the shape of each galaxy's SED while providing coverage for nearly all of the spectroscopic sample, allowing the largest possible sample.
    \item A value must also be chosen for perplexity\footnote{Perplexity sets the Shannon entropy used to determine probabilities in the next step; see \citealt{vanDerMaaten_2008}.} (Figure \ref{fig:perplexity}), one of t-SNE's hyperparameters, which can be interpreted as the number of other galaxies that each galaxy will consider to be its nearest neighbor.  Selection of a low perplexity results in the algorithm focusing primarily on local structure when grouping objects, and selection of a high perplexity results in more global structure (Fig. \ref{fig:perplexity}).  The optimal choice of perplexity will depend upon the catalog size, and unless otherwise noted, the examples shown in this work use a perplexity of 35.
    \item For each pair of objects, t-SNE calculates a probability that one would pick the other as its neighbor, which represents the similarity between the two objects in the initial higher-dimensional space (here seven-dimensional), or equivalently the similarity of the two objects' spectral energy distributions (SEDs).
    \item In the lower-dimensional space (here, two-dimensional was found to be optimal), a point corresponding to each object is first placed randomly, then iteratively moved according to a gradient descent algorithm, making the probability for each pair of objects as similar as possible in the high and low dimensional spaces.  After a set number of iterations, this process ends.  The result is a map where objects with similarly shaped SEDs are grouped together.  The hope is that doing so will simultaneously have grouped objects with similar redshifts and other properties, even though those properties are unknown.
\end{enumerate}

It is evident from the resulting map (Figure \ref{fig:tsnesed}) that grouping objects by SED shape induces an approximate grouping by photometric redshift; most objects with similar colors also have similar $z_{phot}$.  It is also evident that this is not true for all objects; in a small number of cases, $z_{phot}$ differs substantially from that of most close neighbors.  The color of these objects therefore stands out on the map when painted by photometric redshift.

\begin{figure*}[ht]
    \centering
    \includegraphics[width=\textwidth]{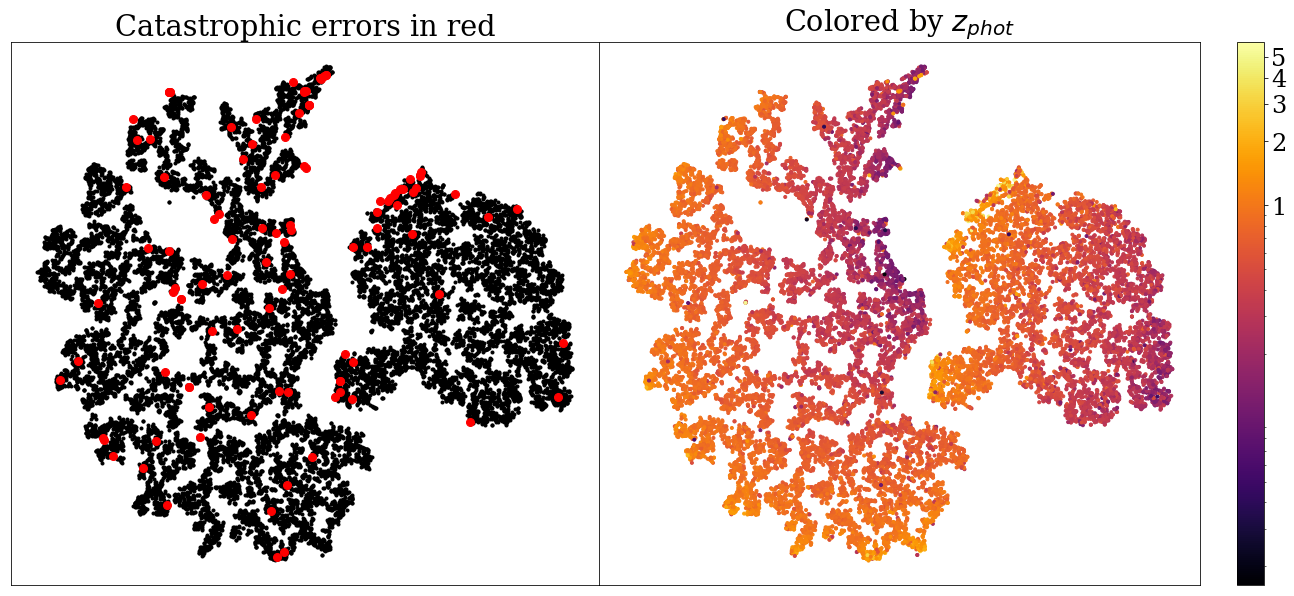}
    \caption{The same t-SNE map (5000 iterations, perplexity 35) colored by $z_{phot}$ (right) and with catastrophic errors in red (left).
    Many of the points where the $z_{phot}$ is different than that of the objects around it (color stands out in plot on right) are also objects with catastrophic $z_{phot}$ errors (red points in plot on left).  Most of these catastrophic errors lie on the edges of clusters in the t-SNE map, implying that their photometry makes them relative outliers when compared with the most similar objects.}
    \label{fig:tsne35}
\end{figure*}

The hope is that these might be the same objects for which spectroscopy reveals that photometric redshifts have been incorrectly determined.  Indeed, an examination of the locations of photometric redshift errors on the resulting t-SNE map shows that many of these outliers have catastrophic errors in $z_{phot}$ (Figure \ref{fig:tsne35}).  Thus, perhaps they are not outliers because they have different redshift than object with similar colors, but instead because their photometric redshifts are incorrect.  

t-SNE is one of several dimensionality-reduction algorithms which could have been selected.  Previously, a self-organizing map (SOM) was similarly shown to be useful for photometric redshift determination \citep{Masters_2015}.  The key difference between the two algorithms lies in the way they fill space.  The SOM spreads out the objects over all available cells, giving every object approximately the same number of neighbors.  In some cases, cells might even be composed of dissimilar objects, which are merely similar relative to other possibilities.  By contrast, t-SNE uses empty space to ensure that nearby objects are truly similar, so that objects with many counterparts will have many neighbors, but objects which are truly outliers will be treated as such.

In the remainder of this section, an algorithm is described to quantify which objects are outliers and attempt to correct their photometric redshifts using information about $z_{phot}$ of their near neighbors.  In \S\ref{sec:results}, the success of that algorithm at identifying and correcting catastrophic errors is evaluated.

\subsection{Comparison with neighbors to determine catastrophic error candidates}

\begin{figure*}[ht]
    \centering
    \includegraphics[width=.98\textwidth]{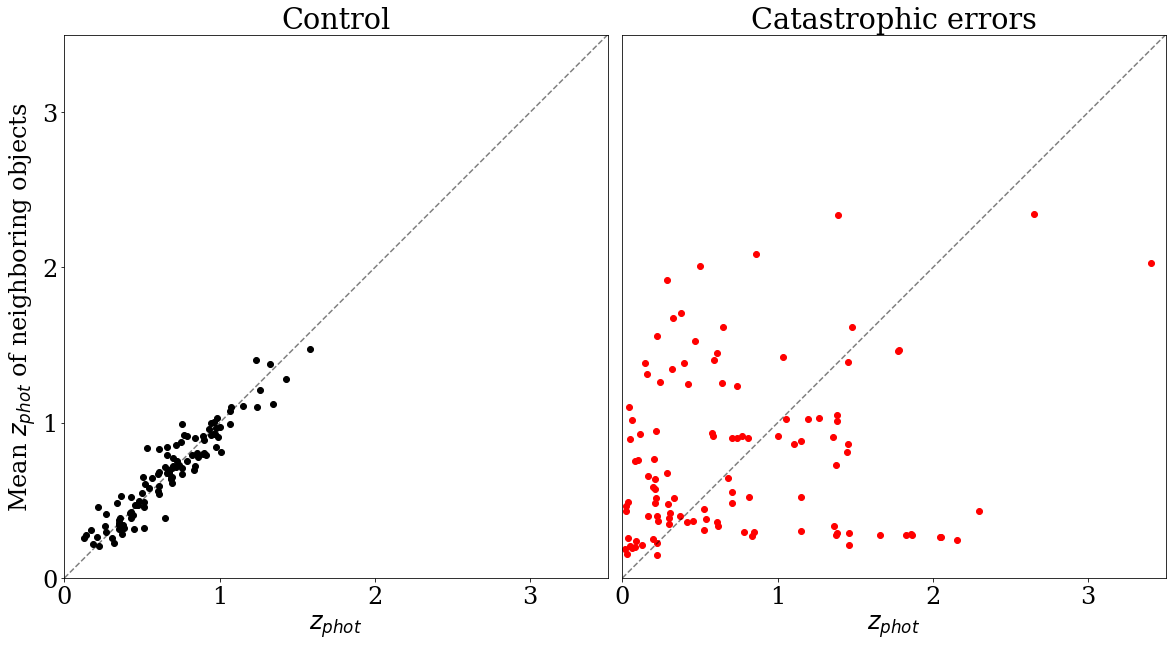}
    \caption{Comparison of $z_{phot}$ of each point with the mean $z_{phot}$ of neighboring objects in t-SNE output, for randomly chosen control sample of 111 objects (left) and for the 111 objects with known catastrophic errors (right). For the random control sample, $z_{phot}$ and the mean neighboring $z_{phot}$ agree very well. For the known catastrophic $z_{phot}$ errors, this is not the case.}
    \label{fig:neighbor}
\end{figure*}

Outliers are selected by taking the mean $z_{phot}$ of all objects within some radius of an object, excluding that object, and comparing this mean with the $z_{phot}$ of the object itself (Figure \ref{fig:neighbor}).  An appropriate radius depends upon the overall density of points in the t-SNE map, attempting to compare the typical object with approximately the same number of neighbors as the choice of perplexity.  For the maps shown in this work, a radius of 5 is selected, so that the typical point on a map with perplexity 35 has approximately 10 neighbors within that radius.

An object is then flagged as an outlier if the difference between its $z_{phot}$ and the mean $z_{phot}$ of its neighbors is greater than 0.5.  This threshold was chosen experimentally in order to optimize the results for the COSMOS catalogs and bands used in this work, and a different threshold may be optimal for other applications.  The choice of threshold is further discussed in \S~\ref{sec:discussion}.  The group of flagged objects is the group predicted to have catastrophic errors in $z_{phot}$.

\subsection{Recalculate outliers} 
\label{subsec:recalc}

For objects flagged as outliers by the method described, the redshifts of objects with similar photometry are then used to produce a new redshift for the flagged object.  Here, two methods are considered, which might repair different sources of catastrophic errors: random measurement errors and the limitations of using a redshift grid of finite precision.

In the first method, a new redshift is produced for each flagged object by replacing its redshift with the mean photometric redshift of its nearest neighbors.  This method would be most effective in a scenario in which an object has photometric measurement errors which have scattered it into a sparse region of color space which EAZY would fit with a very different redshift.  If there are truly two highly-populated, close regions of color space with very different redshifts, then it will be difficult to determine whether an object has been scattered from one to the other.  However, in the fortunate case where photometric errors instead scatter an object into a sparsely populated region, it might be possible to recognize that this is an error and the object was likely scattered from a more highly populated neighboring region, rather than truly having the measured photometry.  Then the redshift of that highly populated region would be the correct one.  Because many of the catastrophic errors lie at the edges of the t-SNE distribution (Fig. \ref{fig:tsne35}, left), this fortunate case may actually be the most common.

In the second method, a new redshift is produced by rerunning EAZY, the template fitting code initially used to determine the photometric redshift of each object and described in \S~\ref{subsec:eazy}, with a significantly finer redshift grid.  In principle, this would be a good idea for every object, not just those flagged, but it would take far too much computational time to run on the full dataset.  This method is considered with two different grids, one 10 times finer and one 100 times finer than the one used initially.

\section{Results} \label{sec:results}

\subsection{Identification of outliers} \label{subsec:identify}

For a control group of objects whose photometric and spectroscopic redshifts are consistent, the mean photometric redshift of nearby objects on the t-SNE map is very frequently the same or very similar.  However, for objects known to have catastrophic errors in photometric redshift determination, the neighboring objects most often have different photometric redshifts than the (incorrectly-determined) photometric redshift of that object.  A threshold is chosen for disagreement between the photometric redshift of an object and its neighbors, and disagreements above that threshold are flagged as catastrophic error candidates.

To quantify the effectiveness of this technique at selecting catastrophic redshift errors at various thresholds, a receiver operating characteristic (ROC) curve is constructed \citep{Albeck1990,Baker2003,Fawcett2006}.  Here, the true positive rate (TPR) is the fraction of catastrophically wrong photometric redshifts which are flagged and the false positive rate (FPR) is the fraction of correct redshifts which are incorrectly flagged.  
\begin{figure}[ht]
    \centering
    \includegraphics[width=.47\textwidth]{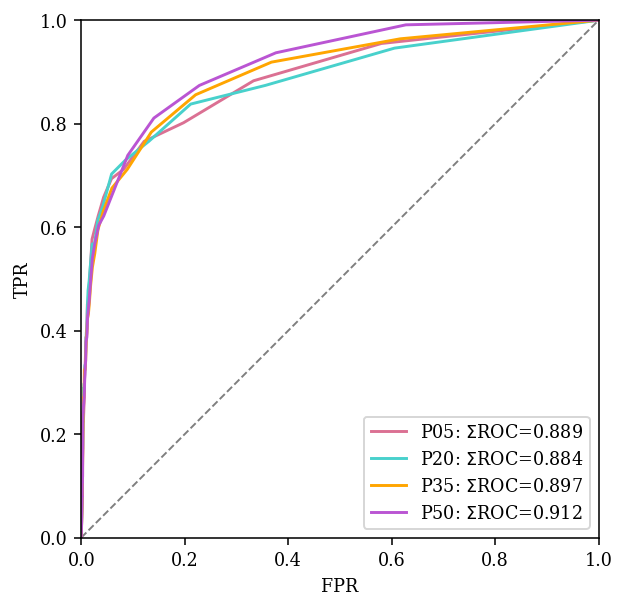}
    \caption{Receiver operating characteristic (ROC) curves for method of flagging an object if its $z_{phot}$ differs from the mean of its neighbors by at least some threshold, for perplexities of 5, 20, 35, and 50.
    For each threshold, we plot the true positive rate at which we correctly flag catastrophic errors against the false positive rate at which we identify correct objects as errors.
    An ROC curve reaching into the top left corner and with a $\Sigma$ROC of 1 would signify a perfect result, while an ROC curve along the diagonal dashed line and with a $\Sigma$ROC of 0.5 would be comparable to guessing randomly.
    Since each curve lies far above the dashed line, this method consistently correctly identifies errors at a much higher rate than it misidentifies correct objects as errors.
    There is no significant variation in the curves for different perplexities, indicating that the method is robust to changes in perplexity.
    Lower values of perplexity perform yield better results for higher thresholds, up to a true positive rate of approximately 0.7, after which higher values of perplexity perform better.}
    \label{fig:roc_p}
\end{figure}

For all thresholds, the true positive rate is higher than the false positive rate, meaning that this method correctly identifies objects with catastrophic errors at a higher rate than it mistakenly flags objects without.  Typically, the choice of any specific threshold will depend upon the relative importance of sample completeness and sample purity.  The integral, or Area Under the Curve (AUC), of the ROC curve $\Sigma ROC$ is an overall quality measure of the flag, equivalent to the probability that a random catastrophic error disagree more with its neighbors than a random non-catastrophic error.  Depending upon the chosen perplexity, $\Sigma ROC$ for this method ranges from 0.884 to 0.912, with the highest value at perplexity 50 for 22,978 total objects (Fig. \ref{fig:roc_p}).  It can therefore be concluded that comparison with neighbors is a useful tool for flagging potential catastrophic redshift errors, without the need for any spectroscopy even as part of a smaller training sample.

\subsection{Repairing outliers} \label{subsec:repair}
The success in identifying catastrophic redshift errors from photometry alone should suggest that it may also be possible to correct them without followup spectroscopy.  Two methods for doing so, described in \S~\ref{subsec:recalc}, are evaluated here.

\begin{figure}
    \centering
    \includegraphics[width=.47\textwidth]{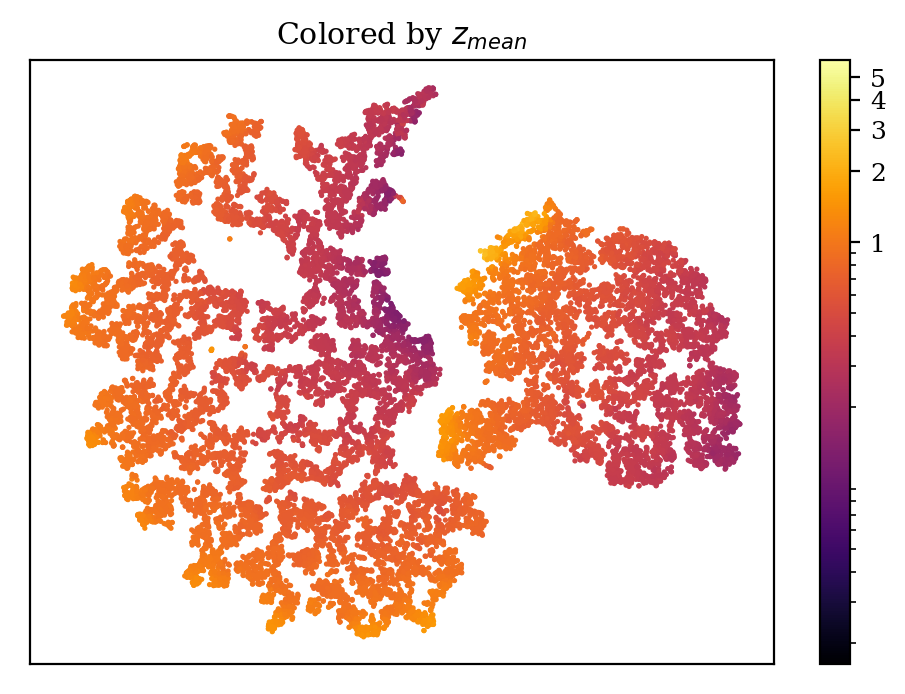}
    \caption{The same t-SNE map of COSMOS catalog with perplexity 35, however here each point is colored by the mean z$_{phot}$ of its nearest neighbors, giving an appearance of smoothness. Note that this map looks similar to the same one colored by z$_{phot}$ (Figure \ref{fig:tsnesed}), but there are no redshift outliers.}
    \label{fig:smoothmap}
\end{figure}

The first method, simply replacing the photometric redshift of every outlier with that of its neighbors, will result in an entirely smooth t-SNE map when colored by this new redshift (Fig. \ref{fig:smoothmap}).  For some objects, this indeed repairs their photometric redshifts, creating agreement with spectroscopy.  However, other objects are not repaired, and ``repairing'' the redshifts of false positives will create a new set of catastrophic errors.  In total, this procedure repaired 23 of the 111 catastrophic errors (63 were not flagged and 25 were flagged but not repaired), while simultaneously creating 308 new catastrophic errors.  It is likely that such a procedure would be more beneficial for a survey with a higher underlying catastrophic error rate; the subset of COSMOS objects with available spectroscopy, which is necessary to evaluate the algorithm, is a brighter and higher-quality sample than the full COSMOS catalog. 

The second method, spending additional computational time to search finer redshift grids, yields similar results.  Searching a finer redshift grid makes EAZY more likely to successfully pick out the minimum $\chi^2$ between templates and photometry, which should in theory improve the accuracy of the determined photometric redshifts.  However, 26 of the 111 catastrophic errors were repaired (as before, 63 were not flagged, so 22 were flagged but not repaired), and 28 new catastrophic errors were created.  As before, the algorithm might repair mistakes and create new ones at similar rates for other catalogs, and thus provide an improvement for a catalog with a higher underlying catastrophic error rate.

A third possible method could have been to look at the full redshift probability density function produced by EAZY and search for additional peaks.  If the primary peak is likely incorrect, perhaps the secondary peak would be the correct redshift solution.  However, most of the 111 flagged objects did not exhibit a second peak.  Of the ones that did, that peak often did not match the spectroscopic redshift.

In summary, t-SNE can be used to successfully flag catastrophic errors at a high rate, allowing the creation of a much higher-purity redshift catalog than current techniques.  However, two attempts to use the same idea to fix these errors, one assuming that the root cause is objects mistakenly scattered into unpopulated regions of photometric color space and the other assuming that the root cause is a failure to find the optimal template match within available computational time, both failed.  Rather, it appears that at present, it is possible to identify likely catastrophic errors without any spectroscopic information, but actually correcting those errors may require spectroscopic followup of likely error candidates.

Perhaps the most likely explanation for this is that many of these catastrophic errors have not merely been scattered randomly into colorspace, but rather represent catastrophic errors in photometry.  For example, an object affected by blending with nearby sources in some bands but not others would produce photometry which does not correctly correspond to any object, rendering attempts to fit it meaningless.  Similarly, a variable object such as an AGN should have photometry which can be well fit at any epoch, but a catalog such as COSMOS2015 is composed of different filters measured over a period of a decade.  Combining those filters would yield a set of photometric measurements which do not correspond to any one epoch, or even to any meaningful object.  

It should also be noted that objects which are saturated in any of the 8 bands used to make the t-SNE map could not be included in the analysis. Thus, the results here are for objects on average fainter and thus more poorly measured than the overall spectroscopic catalog.  These methods would likely perform better on a complete sample.

Support for this idea comes from a closer examination of the location of these catastrophic errors on a t-SNE map (Fig. \ref{fig:tsne35}).  The errors appear in many different locations, indicating that they are not all of a common class merely being misidentified.  However, nearly all of them lie towards the edges of the distribution, indicating that they are relative outliers compared to the bulk of the (well-fit) population.  This would be consistent with photometry which is similar to existing objects in many bands yet has a small number of major photometric errors.  Indeed, 17\% of the objects flagged by t-SNE were also flagged (\texttt{flag\_peter}) as unreliable photometry, compared with 8\% of the overall spectroscopic sample.  If this idea is correct, then the technique described here could become useful not just for identifying catastrophic template fitting errors, but also photometric errors. 

\section{Discussion} \label{sec:discussion}

Although photometric template fitting is successful at finding an accurate redshift for most galaxies in large surveys, for a small number of objects the resulting redshift is wrong by at least 15\%, and in many cases by a factor of 2 or more where templates at different redshifts are nearly degenerate.  Fixing these catastrophic errors is generally only possible with new information, either the inclusion of additional bands which break the degeneracy or spectroscopic followup to determine a far more precise redshift.  Here, a new approach to this problem is considered in an attempt to identify and correct these catastrophic outliers, using the assumptions that most photometric redshifts are correct and that objects with similar photometric colors should lie at similar redshift.  Not only does such an approach not require additional data, but it also avoids the imposition of a luminosity prior, which can bias faint objects towards low redshift and therefore scatter as many as half of all $z > 5$ objects to catastrophically lower redshifts \citep{Steinhardt2014, Steinhardt2016, Davidzon2017}.

The proposed algorithm has four steps: (1) fit photometric redshifts for an entire catalog; (2) use the dimensionality reduction algorithm t-SNE to identify objects with similar photometry; (3) flag objects as catastrophic outlier candidates if their redshift differs significantly from their neighbors' redshifts; and (4) recalculate the redshifts of the flagged objects.  The first three steps are highly successful.  Photometric redshifts are already known to be generally accurate when checked by comparison with spectroscopic redshifts.  The t-SNE map produced shows that even in two dimensions, objects typically have many neighbors, and the photometric redshifts of nearby objects are generally similar.  Finally, a comparison of the spectroscopic and photometric redshifts of flagged catastrophic error candidates demonstrates this method to be highly efficient at selecting outliers with a low false positive rate in the COSMOS2015 catalog.

However, reconstructing the correct redshifts for flagged objects was less successful.  Although an approach relying on neighboring redshifts was able to correct many catastrophic errors, it created a similar number of new errors because of objects which were incorrectly flagged for recalculation.  Similarly, spending additional processing time to use a finer redshift grid for flagged objects produced no net improvement for the COSMOS2015 spectroscopic catalog.  However, this is in part because only $\sim 0.5\%$ of the COSMOS2015 spectroscopic sample exhibits catastrophic errors, and in part because spectroscopic followup is often limited to objects significant brighter than the full survey detection limits.  In a typical catalog, catastrophic error rates might be more than 10 times higher \citep{Hildebrandt_2010, Salvato2019}.  If redshift correction could produce similar performance for these fainter catalogs, something that was not possible to evaluate here, it would provide a substantial improvement in catalog quality.

A more promising interpretation lies in the possibility that these corrections fail because many catastrophic redshift errors might be driven by catastrophic errors in a few of the photometric bands.  Possible causes include not just major measurement errors (which should be relatively unlikely) and saturation (more common), but also blending which only affects lower-resolution measurements and photometry of variable objects built from measurements taken over a range of time.  If this explanation is correct, then the t-SNE approach developed here could be used for future surveys to identify such objects which have escaped identification earlier in the pipeline.

Otherwise, at present, it appears that further improvement requires additional information from new observations.  This raises the possibility of a hybrid approach.  In COSMOS2015, the spectroscopic sample used here included 22,978 objects, or 1.9\% of the full catalog.
Thus, with different target selection, it might be possible to simply take spectroscopic redshifts for the full set of flagged objects, so that the only remaining catastrophic errors would come from the small false negative rate. 

This is a different approach than that used in data-driven photo-z methods.
Usually, spectroscopic information would come first, producing a high-quality training sample on which to then develop estimates for lower-quality data.
Here, we find that, for most applications, existing models and the resulting templates describe photometry well enough that this step is not necessary.
Rather, limited spectroscopic resources can be used to target the objects most likely to need that extra information. Such an approach is likely to become more broadly applicable across a variety of applications in astronomy and other data sciences as {\em a priori} models continue to improve.

\acknowledgments

The authors would like to thank Gabe Brammer, Iary Davidzon, Christian Kragh Jespersen, and John Weaver for useful discussions.  BH thanks Kathryn McEachern for her generous contribution to the SURF program and to this research in memory of David L. Glackin.  CS is supported by ERC grant 648179 "ConTExt".  The Cosmic Dawn Center (DAWN) is funded by the Danish National Research Foundation under grant No. 140. 

Based on data products from observations made with ESO Telescopes at the La Silla Paranal Observatory under ESO programme ID 179.A-2005 and on data products produced by TERAPIX and the Cambridge Astronomy Survey Unit on behalf of the UltraVISTA consortium.




\bibliographystyle{apj}
\bibliography{ref}




\label{lastpage}
\end{document}